\documentclass[%
 aip,
 amsmath,amssymb,
 reprint,%
]{revtex4-1}

\usepackage{graphicx}
\usepackage{dcolumn}
\usepackage{bm}

\usepackage[utf8]{inputenc}
\usepackage[T1]{fontenc}
\usepackage{mathptmx}
\usepackage{etoolbox}
\usepackage[separate-uncertainty = true,exponent-product=\ensuremath{\cdot}]{siunitx}

\makeatletter
\def\@email#1#2{%
 \endgroup
 \patchcmd{\titleblock@produce}
  {\frontmatter@RRAPformat}
  {\frontmatter@RRAPformat{\produce@RRAP{*#1\href{mailto:#2}{#2}}}\frontmatter@RRAPformat}
  {}{}
}%
\makeatother
\begin{document}

\preprint{AIP/123-QED}

\title[High-flux XUV source for coincidence spectroscopy]{Laser-driven high-flux source of coherent quasi-monochromatic extreme ultraviolet radiation for coincidence spectroscopy}
\author{Julian Sp\"athe}
\affiliation{Institute of Optics and Quantum Electronics, Friedrich Schiller University Jena, 07743 Jena, Germany}

\author{Sebastian Hell}
\affiliation{Institute of Optics and Quantum Electronics, Friedrich Schiller University Jena, 07743 Jena, Germany}

\author{Martin Wünsche}
\affiliation{Institute of Optics and Quantum Electronics, Friedrich Schiller University Jena, 07743 Jena, Germany}
\affiliation{Helmholtz Institute Jena, 07743 Jena, Germany}
\affiliation{Indigo Optical Systems GmbH, Moritz-von-Rohr-Str. 1a, 07745 Jena, Germany}

\author{Robert Klas}
\affiliation{Helmholtz Institute Jena, 07743 Jena, Germany}
\affiliation{Institute of Applied Physics, Friedrich Schiller University Jena, 07745 Jena, Germany}
\affiliation{Fraunhofer Institute for Applied Optics and Precision Engineering, 07745 Jena, Germany}

\author{Jan Rothhardt}
\affiliation{Helmholtz Institute Jena, 07743 Jena, Germany}
\affiliation{Institute of Applied Physics, Friedrich Schiller University Jena, 07745 Jena, Germany}
\affiliation{Fraunhofer Institute for Applied Optics and Precision Engineering, 07745 Jena, Germany}

\author{Jens Limpert}
\affiliation{Helmholtz Institute Jena, 07743 Jena, Germany}
\affiliation{Institute of Applied Physics, Friedrich Schiller University Jena, 07745 Jena, Germany}
\affiliation{Fraunhofer Institute for Applied Optics and Precision Engineering, 07745 Jena, Germany}

\author{Thomas Siefke}
\affiliation{Institute of Applied Physics, Friedrich Schiller University Jena, 07745 Jena, Germany}
\affiliation{Fraunhofer Institute for Applied Optics and Precision Engineering, 07745 Jena, Germany}

\author{Gerhard G Paulus}
\affiliation{Institute of Optics and Quantum Electronics, Friedrich Schiller University Jena, 07743 Jena, Germany}
\affiliation{Helmholtz Institute Jena, 07743 Jena, Germany}

\author{Matthias K\"ubel}
\email{matthias.kuebel@uni-jena.de}
\affiliation{Institute of Optics and Quantum Electronics, Friedrich Schiller University Jena, 07743 Jena, Germany}
\affiliation{Helmholtz Institute Jena, 07743 Jena, Germany}

\date{\today}

\begin{abstract}
We present a source of coherent extreme ultraviolet (XUV) radiation with a flux of 10$^{13}$ photons per second at 26.5~eV. The source is based on high-harmonic generation (HHG) in argon and pumped by a frequency-doubled 100~kHz repetition rate fiber laser providing 30~fs pulses centered at 515~nm. We report on the characterization of the source and the generated XUV radiation using optical imaging and photoelectron spectroscopy. The generated radiation is quasi-monochromatized using a suitably coated XUV mirror and used for coincidence spectroscopy of ions and electrons generated from a cold gas target. The high intensity of the focused XUV pulses is confirmed by the observation of two-photon double ionization in argon. Moreover, we demonstrate the capability to perform pump-probe experiments using XUV and visible laser pulses.
\end{abstract}


\maketitle 

\section{Introduction}
The extreme ultraviolet (XUV) spectral region is of great interest to atomic and molecular spectroscopy as the binding energies of valence and inner-valence electrons are situated within the XUV. 
High-harmonic generation (HHG) enables coherent XUV light to be generated through the nonlinear interaction of intense femtosecond laser pulses with gaseous media \cite{Ferray1988, Corkum1993, Lewenstein1994}. The commercial availability of suitable laser sources enable the widespread use of HHG in laboratories around the world.

High-harmonics are uniquely suited for time-resolved experiments, as they are perfectly synchronized with the driving laser field, owing to the underlying generation mechanism. This synchronization has been the key in the development of attosecond science \cite{Krausz2009,CorkumKrausz2007}, which has provided new insights into electron dynamics in atoms \cite{Uiberacker2007, Goulielmakis2010, Ott2013}, molecules \cite{Sansone2010,Calegari2014} and solids \cite{Cavalieri2007,Neppl2015}. Recently, attosecond coincidence experiments have explored the effect of nuclear dynamics on time delays in molecular photoionization \cite{Cattaneo2018,Ahmadi2022}.

The main drawback of HHG has been a relatively low pulse energy. In gas-phase experiments, this leads to a low interaction probability, which has made it difficult to employ high-harmonics as pump pulses. This applies, in particular, to coincidence experiments with their notoriously low signal rates. On this front, significant progress has been made at free-electron laser (FEL) facilities \cite{Rudenko2015}. In contrast to HHG, FELs provide essentially monochromatic femtosecond pulses. With such a clearly defined energy input, it is possible to excite specific electronic states in a molecule, which is advantageous for certain types of experiments \cite{Schnorr2013}.

Previous advances in ultrafast Yb-based laser technology have led to a boost in the average power of HHG radiation \cite{Boullet2009}. Further improvements in the photon flux can be achieved by driving HHG with a visible laser rather than an infrared one \cite{Klas2016,Comby2019}. This results in a significant increase in the photon flux in the $\SIrange{20}{30}{eV}$ range, approaching levels comparable to those of synchrotron sources, while maintaining femtosecond pulse durations \cite{Klas2021}. Moreover, effective monochromatization, i.e., selecting a single harmonic is significantly easier for the large spacing between the harmonics, which is achieved by a short-wavelength driver. 

Here, we describe a high-flux HHG source that is coupled to a reaction microscope for coincidence spectroscopy. We also present first experimental results obtained using this setup. In contrast to previous works \cite{Gagnon2008,Sturm2016}, we use a shorter driving wavelength of \SI{515}{nm} and a significantly higher repetition rate of \SI{100}{kHz}.
The paper is structured as follows: section \ref{section:experimental_setup} describes the experimental set-up consisting of high-harmonic source, diagnostics, beamline and reaction microscope in detail. Section \ref{section:HHG} presents experimental results on the optimization of the HHG efficiency and offers a theoretical description thereof. In section \ref{section:coincidence}, we present results of coincidence experiments using XUV radiation in the reaction microscope. Cross-correlation measurements of XUV and visible pulses are used to deduce the duration of the XUV pulse envelope \cite{Glover1996,Schins1996}. 

\begin{figure*}
\includegraphics[width=\linewidth]{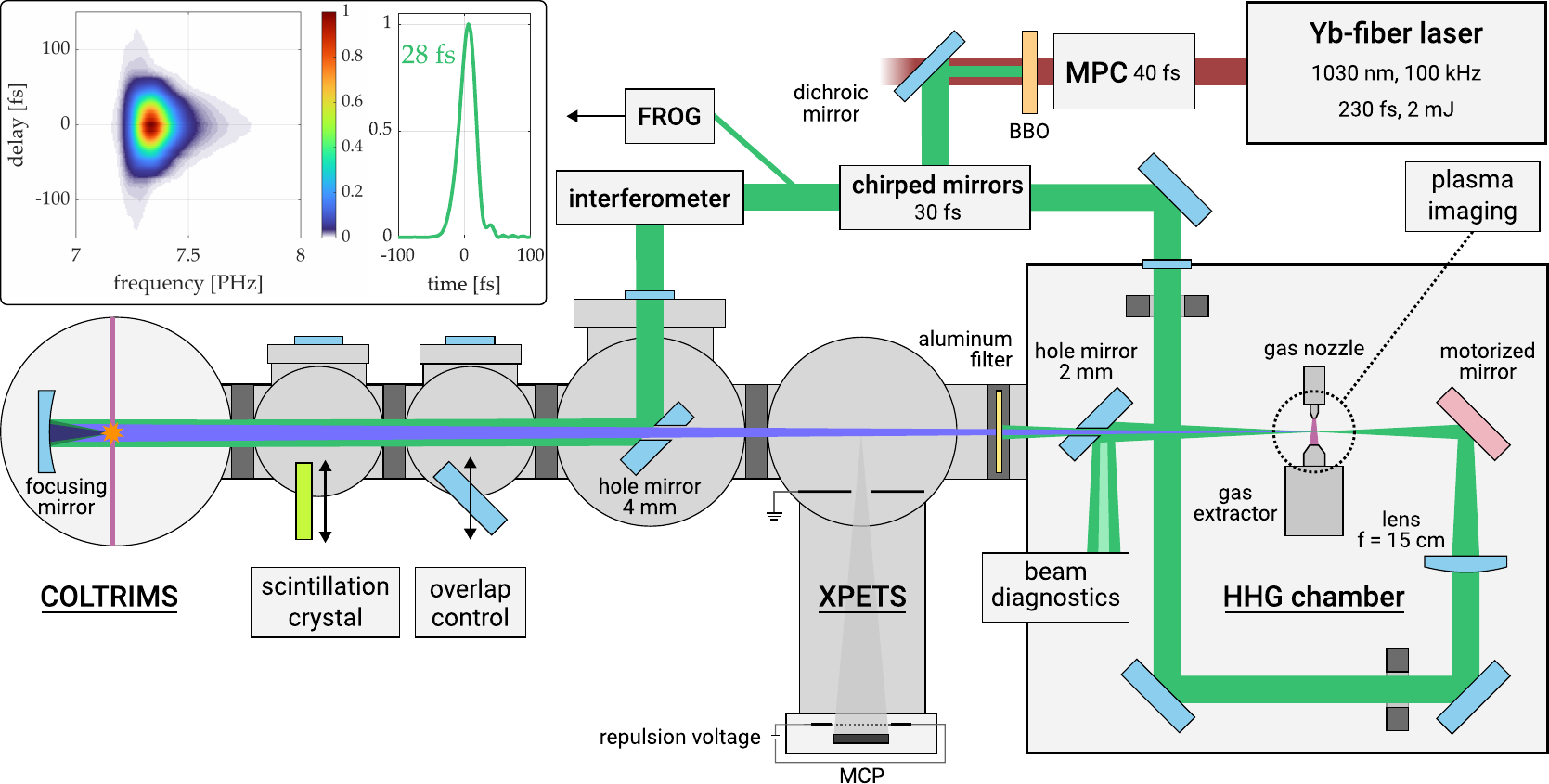}
\caption{\label{fig:xuv_chamber} Overview of the experimental setup consisting of (right to left) HHG chamber, XUV photo electron time-of-flight spectrometer (XPETS), differential pumping stages and the reaction microscope (COLTRIMS). Our laser system consists of a Yb-fiber laser, a multi pass cell pulse compressor, a BBO for frequency doubling and chirped mirrors for visible light pulses. In the HHG chamber, the laser is focused into the argon gas target using a thin lens (\SI{150}{mm} or \SI{200}{mm} focal length). The plasma generated in the laser focus is imaged onto a camera outside the chamber. A hole mirror followed by a thin aluminum filter are used to subtract the laser from the generated XUV beam. Further, the reflected laser is used for beam diagnostics. The XPETS is used for live measurements of the XUV spectrum and photon number estimation. In order to reduce the pressure from \SI{e-3}{mbar} to $\leq\SI{e-10}{mbar}$ in the COLTRIMS, the XUV beam propagates through three differential pumping stages (DPS). In addition, the second DPS is used to recombine the XUV beam with the \SI{515}{nm} beam for two-color experiments. The overlap of the two beams can be checked by moving in a mirror in the third DPS. A scintillation crystal is used to image the XUV beam. Finally, a mirror inside the COLTRIMS with Si-Sc coating focuses ($f = \SI{75}{mm}$) the XUV beam on the cold gas jet target. The inset shows a typical FROG (Frequency Resolved Optical Gating) trace and the retrieved pulse duration of \SI{28}{fs}, with additional dispersion up to the focus taken into account.}
\end{figure*}

\section{Experimental setup}
\label{section:experimental_setup}
\subsection{Laser system}

We use a ytterbium-doped fiber laser (Active Fiber Systems, AFS), which provides up to \SI{200}{W} of average power at \SI{1030}{nm} with \SI{100}{kHz} repetition rate and a pulse duration of \SI{230}{fs}. A multipass cell (AFS) filled with $\approx \SI{1.3}{bar}$ of argon as non-linear medium is used for pulse compression below \SI{40}{fs} at $>$\SI{85}{\percent} efficiency.\\
The post-compressed \SI{1030}{nm} laser pulses are frequency-doubled in a beta barium borate (BBO) crystal of \SI{300}{\micro\meter} thickness. We generate the second harmonic at \SI{515}{nm} with an efficiency of $\approx \SI{20}{\percent}$. Positive chirp due to temporal walk-off in the BBO is compensated using chirped mirrors.\\
The choice of \SI{515}{nm} to drive the HHG is based on the single-atomic scaling law\cite{Colosimo2008, Klas2021a, Shiner2009} for the intensity of the harmonic order $q$,  $I_q \propto \lambda^{-5\dots -7}$, which predicts up to two orders of magnitude higher HHG efficiency compared to \SI{1030}{nm}.

\subsection{HHG chamber}
The XUV source (Fig.\ref{fig:xuv_chamber}) is build inside a vacuum chamber with a footprint of \SI{550}{mm}$\times$\SI{490}{mm} and a height of \SI{350}{mm}. The laser is focused by a thin lens ($f = 150\,$mm or $f = 200\,$mm) into an argon gas jet in order to generate XUV radiation by HHG. The last mirror before the focus defines the position of the XUV beam in the lab frame. Behind the focus, a perforated mirror with a \SI{2}{mm} hole drilled at \SI{45}{\degree} is utilized to reflect $\approx \SI{90}{\percent}$ of the driving laser beam, whilst permitting the XUV beam to pass through. As a second step of separation, an aluminum filter (\SI{150}{nm}) blocks the residual driving laser while transmitting up to \SI{35}{\percent} of the XUV radiation. The transmittance was measured using the XPETS.\\
The gas target for HHG in the form of an argon gas jet is provided by a nozzle of \SI{300}{\micro\meter} diameter. The position of the gas jet in relation to the laser focus can be adjusted in three dimensions by motorized stages. As discussed later, this is crucial to optimize the phase matching conditions. 
A second movable extractor nozzle (\SI{1}{mm} diameter) directly opposite the gas nozzle and connected to a separate scroll pump (\SI{15}{\meter\cubed/\hour}) extracts most of the argon gas from the chamber.
At a backing pressure of \SI{1}{bar}, the HHG chamber is kept at a pressure of $\approx \SI{e-3}{mbar}$. This is essential to minimize reabsorption of the generated XUV radiation.

\subsection{Plasma diagnostics}

\begin{figure}
\includegraphics[width=\linewidth]{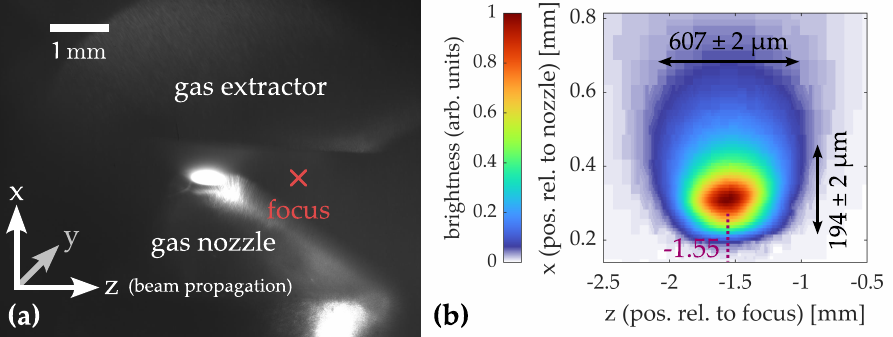}
\caption{\label{fig:plasma_diagnostics} Plasma image and evaluation. In (a) an image taken with the plasma imaging setup shows the plasma in the argon gas jet. Both, the gas nozzle and extractor are visible. The beam propagates along the z-axis and the gas jet along the x-axis. In (b) the dimensions (full width at half maximum, FWHM) and position of the plasma relative to laser focus (z-axis) and gas nozzle (x-axis) are determined.}
\end{figure}

To navigate the gas and extractor nozzles relative to the laser focus, it is helpful to monitor the laser-generated plasma in the HHG gas jet. For this purpose, a set of lenses is used to image the plasma in top view to a camera outside the chamber (see Fig.~\ref{fig:plasma_diagnostics} (a)).
Thus, the plasma position under experimental conditions, i.e., in an argon gas jet and high laser intensity, can be determined relative to the laser focus. Using the known nozzle size as a reference further allows us to measure the plasma dimensions. For the case shown in Fig.~\ref{fig:plasma_diagnostics} (b), the dimensions of the plasma are $\SI{607+-2}{\micro\meter}$ along the laser propagation and $\SI{194+-2}{\micro\meter}$ along the gas jet.\\
The calibration procedure proposed by \textit{Comby et al.} \cite{Comby2018a} enables us to determine the gas pressure of the jet as a function of the plasma brightness. Briefly, the high-harmonic chamber is filled with a static pressure of argon and the plasma brightness is measured as a function of the pressure. We determined an absolute argon pressure of $\approx\SI{340}{mbar}$ for the case of Fig.~\ref{fig:plasma_diagnostics}.

\subsection{XUV beam imaging}
\begin{figure}
\includegraphics[width=\linewidth]{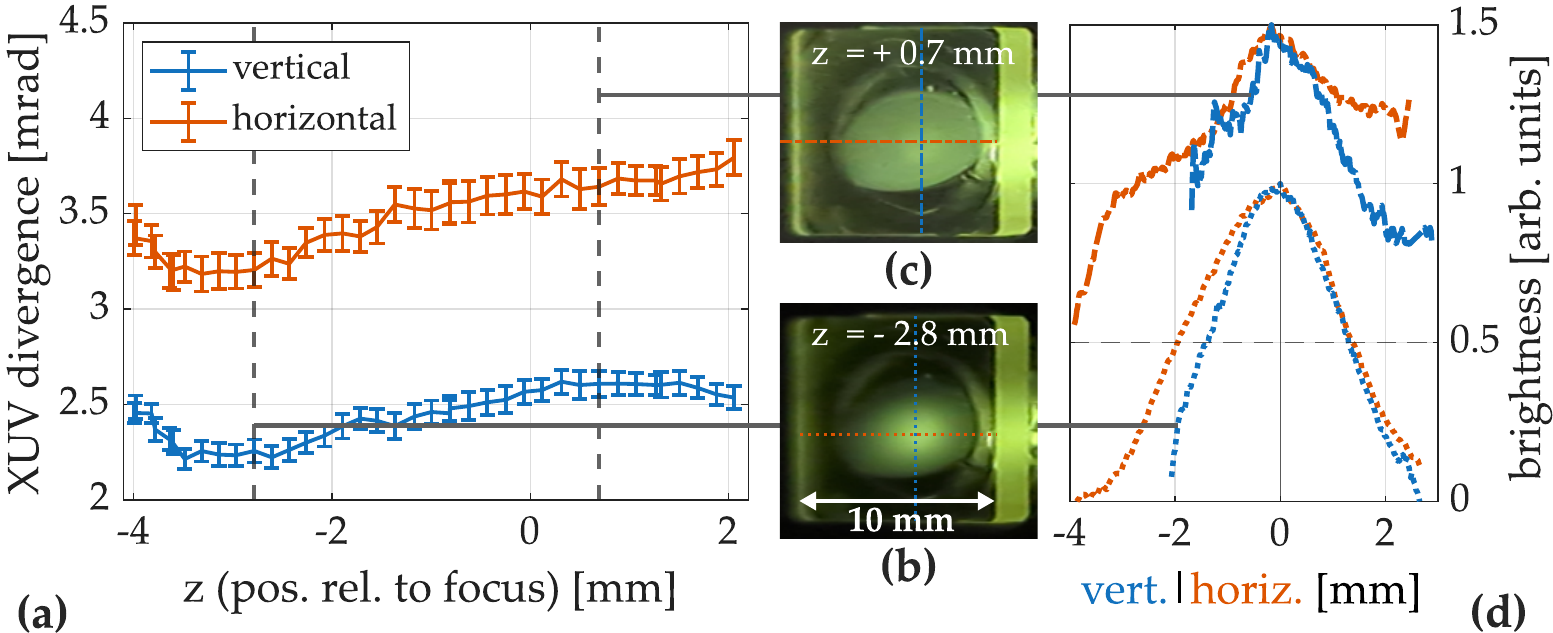}
\caption{\label{fig:scinty} XUV divergence determined by scintillation imaging. In (a), the vertical and horizontal half angle divergence ($1/e^2$) of the XUV beam in dependence on the XUV generation position (z-axis) are shown. The images of the scintillation crystal hit by XUV radiation generated before (b) and after the focus (c) are presented as an example for the increasing divergence. In the latter case, the beam is cut due to apertures in the beam path. The round feature around the beam is caused by the crystal support structure. The maximum brightness divided by exposure time of (b) is 4.5 times higher compared for (c). Panel (d) shows the horizontal and vertical lineouts for (b) and (c) with normalized brightness for comparison.}
\end{figure}

The XUV beam is imaged between HHG chamber and COLTRIMS (Fig.\ref{fig:xuv_chamber}) using a scintillation crystal (YAG:Ce, $\SI{10}{mm}\times\SI{10}{mm}\times\SI{0.5}{mm}$). In typical conditions, the XUV flux is sufficient to see the scintillation with bare eye. The beam image enables an accurate alignment of the XUV beam's pointing and allows determining its profile and divergence. Fig.~\ref{fig:scinty} (a) shows the measured vertical and horizontal divergence of the XUV beam, which depends on the XUV generation position, i.e., the position of the nozzle relative to the laser focus. The smallest divergence is obtained when the gas nozzle is placed in front of the focus position. This finding agrees with the predictions made for short trajectories in Ref.~\cite{Wikmark2019}. The small divergence allows the beam to pass a series of apertures within the beam path for differential pumping as well as the perforated mirror (see Fig.~\ref{fig:scinty} (b)). A higher divergence can result in the beam being cut (see Fig.~\ref{fig:scinty} (c)).  For a quantitative comparison of the beam profiles (b) and (c), their horizontal and vertical lineouts are shown in Fig.~\ref{fig:scinty} (d).

\subsection{Live XUV spectroscopy and estimation of the photon flux}
\begin{figure}
\includegraphics[width=\linewidth]{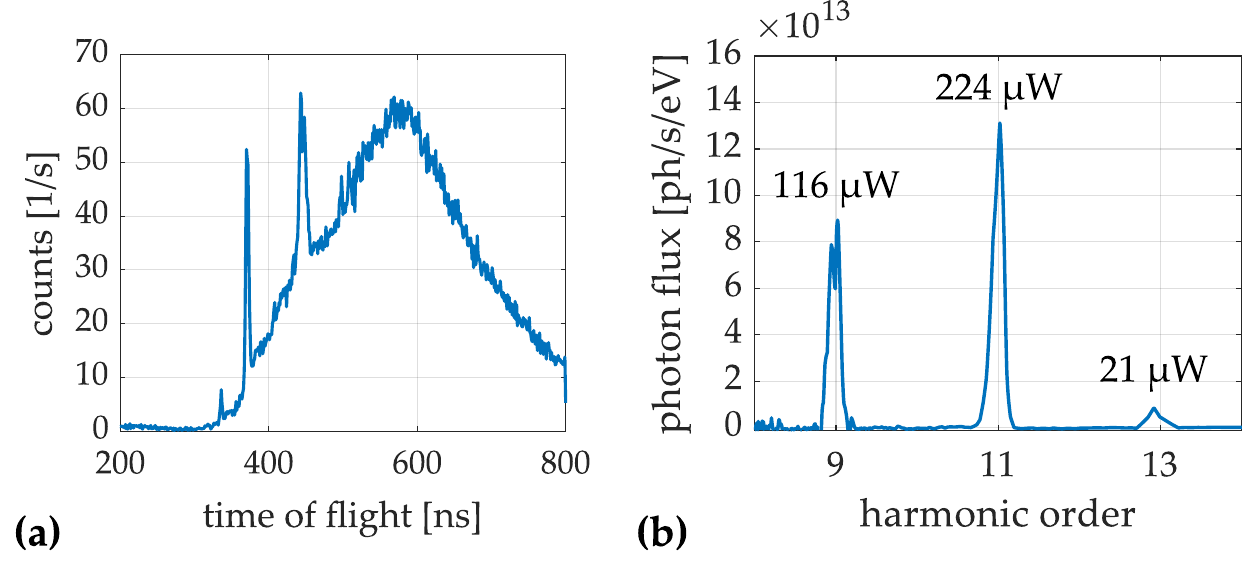}
\caption{\label{fig:XPETS_setup} XPETS spectrum. In (a), the time-of-flight photoelectron spectrum captured with the XPETS is presented. It is dominated by the prominent harmonic peaks on top of a broad background due to scattered electrons. The post-processed photon energy spectrum is shown in (b). An estimation of the photon number per second allows determining the average power of the harmonic lines available in the COLTRIMS (i.e, behind the aluminum filter) with a precision of $\approx\SI{75}{\percent}$. The measurement was carried out using a focusing lens of $f = \SI{200}{mm}$, position relative to focus $z = \SI{-0.8+-0.1}{mm}$, laser power $P_L = \SI{5.3\pm 0.1}{W}$ after iris, and argon backing pressure of $p = \SI{1.15+-0.05}{bar}$.}
\end{figure}

In order to measure the spectrum of the generated XUV radiation, we employ a photoelectron time-of-flight spectrometer (XPETS) connected to the first differential pumping stage (DPS) following the HHG chamber (see Fig.\ref{fig:xuv_chamber}). The XPETS consists of a \SI{350}{mm} long flight tube with $\mu-$metal shielding, and a microchannel plate (MCP) detector at its end. The first DPS is filled with approximately $10^{-5}$ mbar of argon. Ionization of the argon atoms by the propagating XUV leads to a count rate of the order of several 100 electrons per second on the MCP detector. This is sufficiently high to optimize the HHG generation conditions. A mesh on a negative potential (repulsion voltage $U_{rep}$) immediately before the MCP is used to suppress the low-energy background due to scattered electrons.\\
The TOF spectrum (Fig.~\ref{fig:XPETS_setup}~(a)) clearly shows sharp lines corresponding to the high-order harmonics in the XUV spectrum. To assign the correct harmonic order, $|U_{rep}|$ is increased until a harmonic line vanishes. The potential then corresponds to the kinetic energy of the photoelectron, $E_{\text{kin}}$. It is connected to the photon energy $E_{\text{ph}}$ via the ionization potential $E_{\text{IP}}$ of argon:
\begin{equation}
    E_{\text{kin}} = E_{\text{ph}} - E_{\text{IP}},
\end{equation}
with $E_{\text{IP}}$ = \SI{15.760}{eV}\cite{Velchev1999}. 

The XPETS signal can be used to estimate the photon flux (Fig.~\ref{fig:XPETS_setup}~(b)). To this end, the measured photoelectron count rate $N_{\text{el}}$ is corrected for the continuous background. $N_{\text{el}}$ connects to the number of photons $N_{\text{ph}}$ via the photo-ionization cross-section of argon\cite{Samson2002} $\sigma_{\text{Ar}}(E_{\text{ph}})$, the number density of argon atoms $n_{\text{Ar}} = p_{Ar}/k_BT $ and the efficiency to capture electrons $\eta(E_{\text{ph}})$:
\begin{equation}
    \label{eq:Nel}
    N_{\text{el}} = N_{\text{ph}} \cdot \sigma_{\text{Ar}}(E_{\text{ph}}) \cdot n_{\text{Ar}} \cdot \eta(E_{\text{ph}}).
\end{equation}
For our setup, we have $\eta(E_{\text{ph}}) = \eta_{\text{MCP}} \cdot \eta_{\text{spec}}(E_{\text{ph}})$ with MCP efficiency $\eta_{\text{MCP}} = \SI{0.4+-0.2}{}$ and spectrometer efficiency $\eta_{\text{spec}}(E_{\text{ph}})$, which describes the fraction of emitted photoelectrons $\alpha (x,E_{\text{ph}})$ that hit the detector integrated over the ionization length covered by the detector $l_i = x_2 - x_1$:
\begin{equation}
    \eta_{\text{spec}}(E_{\text{ph}}) = \int_{x_1}^{x_2}~\text{d}x~\alpha(x,E_{\text{ph}}) .
\end{equation}
Fraction $\alpha$ and length $l_i$ are determined by a particle tracing simulation based on the spectrometer geometry and the local magnetic field at the interaction region. For the 11th harmonic (H11, \SI{26.5}{eV}, \SI{46.8}{nm}) $\eta_{\text{spec}} = 8.0^{+2.6}_{-0.6}\cdot10^{-9}$.\\
To calculate $N_{\text{ph}}$, we rearrange Eq.\ref{eq:Nel} to
\begin{equation}
    N_{\text{ph}} = N_{\text{el}} \cdot \frac{k_B T}{p_{\text{Ar}} \cdot \sigma_{\text{Ar}}(E_{\text{ph}}) \cdot \eta_{\text{MCP}} \cdot \eta_{\text{spec}}(E_{\text{ph}})},
\end{equation}
and consider temperature $T = \SI{294}{K}$ and argon gas pressure $p_{\text{Ar}} = \SI{2.75+-0.14e-5}{mbar}$. $k_B$ is the Boltzmann constant.\\
From the count rate of \num{411} photoelectrons per second for the 11th harmonic (H11, \SI{26.5}{eV}, \SI{46.8}{nm}), we estimate a photon flux of $5.3^{+4.6}_{-3.3}\cdot10^{13}~\text{photons/s}$ which corresponds to $224^{+195}_{-134}~\text{µW}$ behind the aluminum filter. This is equivalent to a conversion efficiency from \SI{515}{nm} to H11 of $1.2^{+1.1}_{-0.7}\cdot10^{-4}$ before the aluminum filter. As the uncertainties indicate, this is a rough estimate of the XUV flux and is intended only as a measure of the order of magnitude of the flux of the presented XUV source.\\
The described XUV photoelectron time-of-flight spectrometer (XPETS) allows measuring XUV spectra and, at the same time, using the full beam for the primary experiment in the reaction microscope.

\subsection{XUV focusing and quasi-monochromatization}

In order to conduct coincidence measurements, the XUV beam is separated from the generating beam and propagates into the cold target recoil ion momentum spectrometer (COLTRIMS)\cite{Ullrich2003}, also known as reaction microscope. The COLTRIMS is able to detect electrons and ions in coincidence. Since the measurement of clean coincidences, requires a pressure of $\leq \SI{1e-10}{mbar}$, we use several differential pumping stages between HHG chamber ($\approx \SI{e-3}{mbar}$ mbar) and COLTRIMS (Fig.~\ref{fig:xuv_chamber}). The XUV beam is focused onto the cold target jet by a scandium silicon multi-layer mirror \cite{Yulin2004} with 75 mm focal length, which reflects \SI{40}{\percent} of H11 and $<\SI{5}{\percent}$ for other harmonics\footnote{The reflectivity values are manufacturer specifications.}. This results in a quasi-monochromatic XUV pulse (see Fig.~\ref{fig:XUV515} (b)). For two-color pump-probe experiments using both XUV and intense visible or infrared light, the XUV mirror is additionally coated with \SI{28}{nm} aluminum. This affects the XUV reflectivity only slightly while increasing the reflectivity in the visible and infrared to more than \SI{90}{\percent}, and obtain intensities in excess of \SI{1e14}{W/cm^2} in the laser focus.

\section{Optimization of HHG efficiency}
\label{section:HHG}

Maximizing the efficiency of the HHG process requires a look at the response of the medium to the laser field. The simplified\footnote{The initial inhomogeneous wave equation is evaluated for plane wave propagation, paraxial and slowly varying envelope approximation as well as in a 1D model without reabsorption.} theoretical expression for the intensity $I_q$ of the harmonic line with order $q$ is \cite{Weissenbilder2023,Klas2021a,Chang2011}
\begin{equation}
    \label{eq:Iq}
    I_q \propto \varrho^2 |d_q|^2 L^2 \text{sinc}^2\left(\frac{\Delta k_q L}{2}\right),
\end{equation}
with the density of the gas medium $\varrho$ and gas medium length $L$. The single atom dipole amplitude $d_q$ is the microscopic response of the gas medium to the laser field and depends on the laser intensity in a highly nonlinear fashion ($|d_q|^2 \propto I_{\text{laser}}^{4.6}$ \cite{Lewenstein1994,Klas2021a}). The term $\Delta k_q$ represents the wave vector mismatch between the initial laser field and the generated harmonic field of order $q$, i.e, it represents the macroscopic response. A vanishing mismatch $\Delta k_q$ corresponds to perfect phase matching and maximizes the harmonic intensity $I_q$.\\
We can relate these theoretical parameters to the parameters available in the experiment. The gas density is easily adjusted by the backing pressure of the gas jet. The medium length can be tuned by shifting the gas nozzle on the x-axis. Both are covered in detail in Ref.~\cite{Weissenbilder2022}. The peak intensity $I_0 = I(z=0)$ is changed by adjusting the laser power with half-wave plate and polarizer. By varying the beam size with an iris, the focusing geometry and consequently the intensity profile $I(z,r)$ is changed. Shifting the gas nozzle relative to the laser focus along the z-axis will change the intensity $I(z)$ at jet position $z$. The influence of the two latter parameters is investigated in more detail in this paper. A similar investigation for a near infrared driving laser is shown in Ref.\cite{Major2021}

\subsection{Experimental parameter scans}

\begin{figure}
\includegraphics[width=\linewidth]{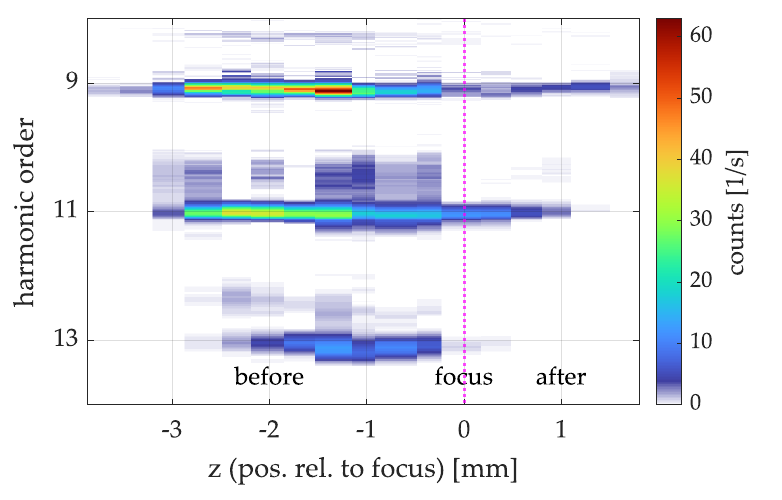}
\caption{\label{fig:z_scan} Scan of the HHG position. The nozzle  position relative to the laser focus is scanned from \SI{-3.5}{mm} to \SI{1.8}{mm}. The laser power was $P_L$ = \SI{7.5}{W} with the iris opened, at an argon pressure of p = \SI{2}{bar}.}
\end{figure}

To scan the generation position of the high harmonics relative to the laser focus, the gas nozzle is shifted along the beam propagation axis (z-axis) with all other parameters fixed (\SI{2}{bar} argon backing pressure, \SI{7.5}{W} laser power, optimized for maximum flux) and fully opened iris (beam radius \SI{4}{mm}). Fig.~\ref{fig:z_scan} shows the resulting spectra for a scan from approximately \SI{-3.5}{mm} (before focus) to \SI{1.8}{mm} (after focus) in steps of \SI{0.4}{mm}. For all harmonic orders, the generation before the nominal focus position is more efficient with maximum yield at $z = \SI{-1.16}{mm}$ (H9, H13) and $z = \SI{-2.19}{mm}$ (H11), respectively. The Rayleigh length assuming a Gaussian pulse with $M^2 = 1.2$ is \SI{277}{\micro\meter} (cf. Appendix). Behind the focus, the HHG signal nearly vanishes. This is in contrast to several earlier publications, e.g. Ref.~\cite{Salieres1995}. However, some recent work also reports on HHG before the focus \cite{Kretschmar2024}.

\begin{figure}
\includegraphics[width=\linewidth]{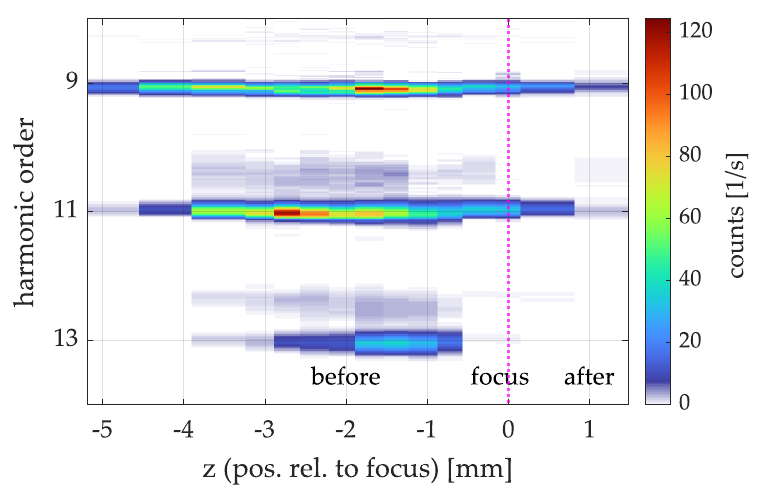}
\caption{\label{fig:zo_scan} Optimization scan. The generation position of the XUV radiation relative to the focus is scanned from \SI{-4.6}{mm} to \SI{1.5}{mm}. In addition, the iris is optimized for each step. Further parameters: laser power $P_L = \SI{7.0}{W}$ before iris, argon pressure $p = \SI{2}{bar}$.}
\end{figure}

In a second scan, at each nozzle position $z$ the iris diameter was optimized to obtain maximum XUV flux. Closing the iris corresponds to decreasing the beam diameter and the laser power. However, the resulting spectra in Fig.~\ref{fig:zo_scan} show an increase of the yield compared to an open iris. The maxima of the yield are again located before the focus for all harmonics. Compared to the measurement with opened iris (Fig.~\ref{fig:z_scan}), the yield for H11 is three times higher and the maximum slightly further away from the focus ($z = \SI{-2.58}{mm}$). For this position, the beam radius (laser power) is reduced to \SI{3.4}{mm}, i.e., \SI{85}{\percent} of the full diameter.

\subsection{Discussion of the phase matching conditions}
An explanation of the observed behavior of the HHG efficiency regarding the generation position requires consideration of the four contributions to the wave vector mismatch $\Delta k$: dispersion in neutral atoms ($\Delta k_\mathrm{at}$), dispersion in plasma due to free electrons ($\Delta k_\mathrm{fe}$), phase variations due to laser focusing ($\Delta k_\mathrm{foc}$) and the intrinsic phase as result of the microscopic single atom response ($\Delta k_{i}$).~\cite{Weissenbilder2022}\\
First, we can define the critical ionization degree as \cite{Rundquist1998,Weissenbilder2022}
\begin{equation}
    \eta_\mathrm{crit} = \frac{\Delta k_\mathrm{at}}{\Delta k_\mathrm{fe}} \eta.
\end{equation}
In general, the ionization degree $\eta$ quantifies the fraction of atoms ionized when the main peak of the laser field arrives at the target. The critical ionization degree $\eta_\mathrm{crit}$ is the upper limit of $\eta$ for perfect on-axis phase matching \cite{Weissenbilder2023}. This leads directly to an upper limit for $I_0$. For the experimental parameters and H11 this yields $\eta_{crit} = 0.16$, which corresponds to $I_{0,\mathrm{crit}} = \SI{2.5e14}{W\per cm^2}$. The estimated peak intensity in the experiment, $I_0\approx \SI{1.7e15}{W\per cm^2}$, is much higher. The high ionization degree corresponds to a large number of free electrons, causing a high plasma dispersion $\Delta k_\mathrm{fe}$ which cannot be compensated by the other contributions. Thus, HHG is more favorable at lower intensity, i.e., away from the focus.

\begin{figure}
\includegraphics[width=\linewidth]{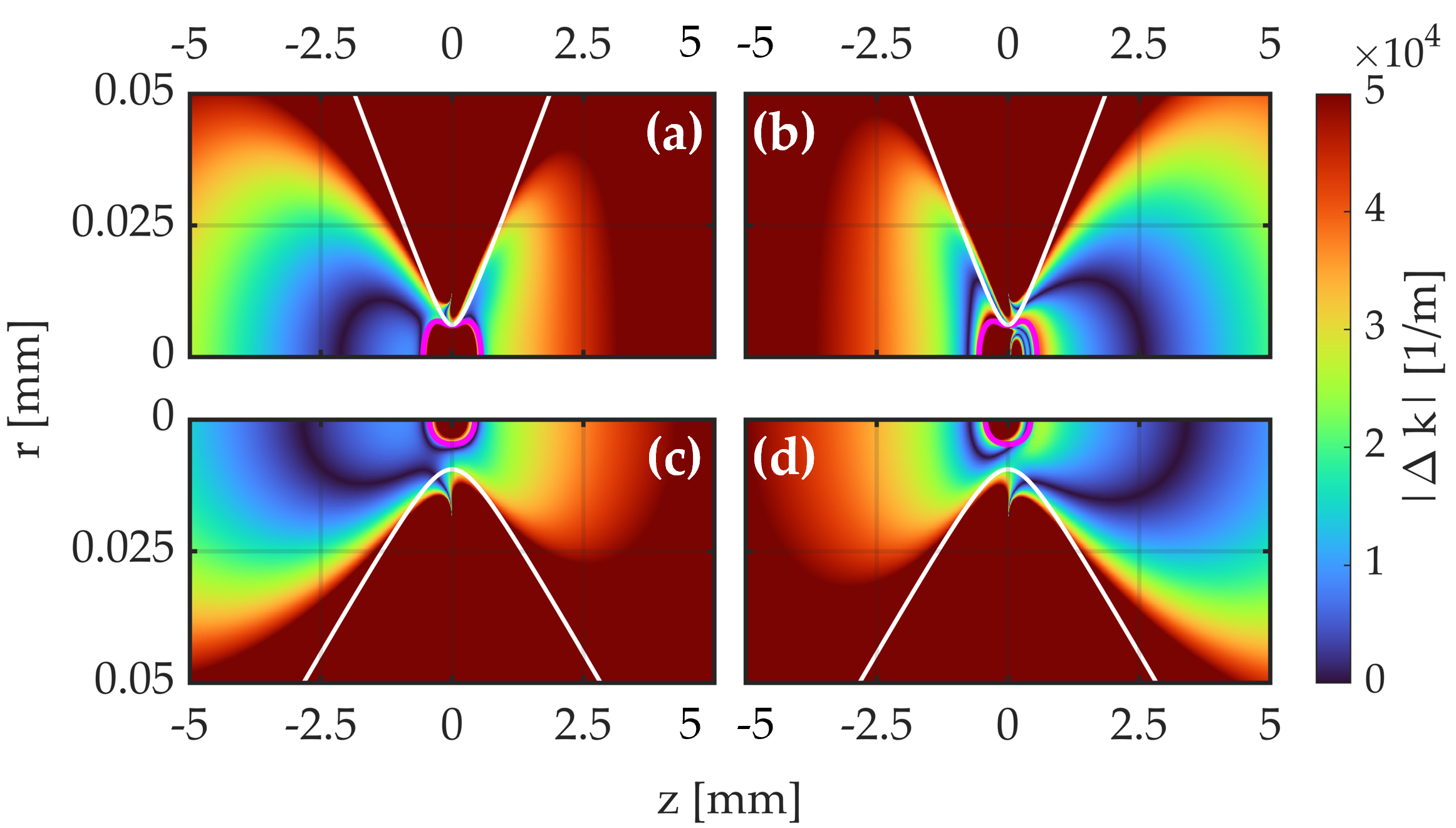}
\caption{\label{fig:pmm} Phase matching maps for Harmonic order 11. The absolute wave vector mismatch $|\Delta k|$ as a measure of phase matching is shown for the focus area for short (a,c) and long (b,d) trajectories as well as for fully opened (a,b) and partially closed iris (c,d). The calculations are preformed for the parameters used in the experiment: laser power $P_L = \SI{7.5}{W}$, peak intensity for opened iris ($w = \SI{4}{mm}$) $\SI{1.7e15}{W\per cm^2}$ and with partially closed iris ($w = \SI{2.6}{mm}$) $\SI{4.1e14}{W\per cm^2}$; absolute argon pressure \SI{337}{mbar}).}
\end{figure}

To answer why we achieve more efficient HHG with the nozzle in front of the focus, we calculated the whole wave vector mismatch $\Delta k$ for the conditions of our experiment. An approximated analytical calculation of the intrinsic phase, proposed by Weissenbilder \textit{et al.} in Ref.~\cite{Weissenbilder2022,Weissenbilder2023}, is used. Fig.~\ref{fig:pmm} shows the computational result for $|\Delta k|$ in the focus area. Dark blue areas with $\Delta k \approx 0$ are favorable for HHG. For a high efficiency, good phase matching conditions over a wide range are required. Fig.~\ref{fig:pmm} (a) shows that this is fulfilled by the short trajectories around $z = \SI{-2.0}{mm}$, i.e., before the focus. This is in good agreement with the experimental observations. For long trajectories (b), good phase matching is achieved around $z = \SI{+2.5}{mm}$. 
However, as seen in Fig.~\ref{fig:scinty}), the divergence is much larger when harmonics are generated behind the focus, which will lead to clipping of the XUV beam within our setup. Moreover, the distance from the focus is slightly larger, leading to somewhat lower intensity. This may explain the low XUV flux measured in the XPETS for HHG generation behind the focus.

The enhancement effect of partly closing the iris can be explained by the change of the focusing geometry. Closing the iris reduces the beam diameter, which results in an increased Rayleigh length and beam waist. Since the iris cuts off parts of the beam, the laser power and thus the peak intensity $I_0$ is decreased. However, the larger focal area leads to an increase of the Rayleigh length, and hence the intensity $I(z)$ at the generation position may in fact increase, corresponding to larger microscopic response.  In addition, the phase matching conditions stay approximately the same or may even be improved, cf.~Fig.\ref{fig:pmm} (a) and (c). 

By iteratively optimizing all parameters, i.e., first, setting the gas pressure; second, optimizing  generation position z; third, adjusting the iris size, we optimize the HHG yield and obtain an estimated flux of \SI{5e13}{photons/s} at H11 at $z = \SI{-2.6}{mm}$.

\section{Coincidence measurements with XUV radiation}
\label{section:coincidence}

\subsection{Non-linear photoionization with XUV}

\begin{figure}
\includegraphics[width=\linewidth]{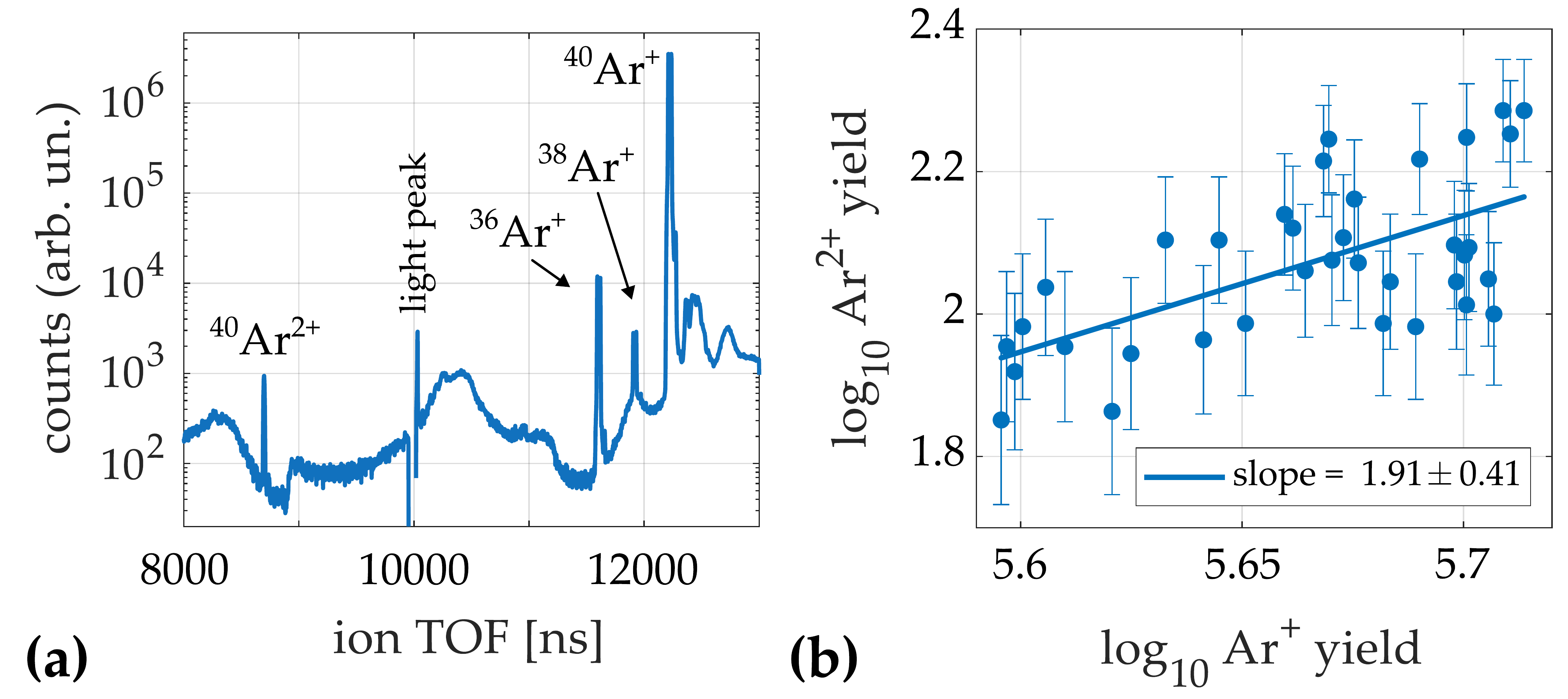}
\caption{\label{fig:Arpp} XUV photoionization experiment in argon. In (a), the ion time-of-flight (TOF) spectrum shows the signal corresponding to the three isotopes of Ar$^+$ and a weak signal of Ar$^{2+}$, as indicated. In (b), the yields of Ar$^+$ and Ar$^{2+}$ are plotted for time spans of \SI{30}{min} to retrieve the non-linearity of the double ionization process.}
\end{figure}

For a first coincidence measurements, a concave ($f = \SI{75}{mm}$) Si-Sc multilayer mirror, optimized for the reflection of H11, is used to focus the XUV radiation into a cold gas jet of argon atoms inside the reaction microscope. In Fig.~\ref{fig:Arpp} (a), we present the recorded time-of-flight (TOF) spectrum for the ions from the target jet. Besides a strong signal from Ar$^+$ at \SI{12200}{ns} at a rate of \SI{390}{Hz}, we also observe a peak at \SI{8700}{ns} at a rate of \SI{0.1}{Hz}, corresponding to the mass-over-charge ratio of \SI{20}{a.m.u / e}. This signal may correspond to Ar$^{2+}$, which can only be produced by two-photon absorption at the present photon energy. However, the signal could also originate from a small contamination of Ne in our gas source.

In order to show that the signal at \SI{8700}{ns} is indeed due to two-photon double ionization of Ar, we assess the non-linearity of the signal. The ratio of the total counts amounts to $\text{Ar}^{2+}/\text{Ar}^+ = \num{1.8e-4}$. This ratio is evaluated for time slices of \SI{30}{min} throughout the measurement, as shown in Fig.~\ref{fig:Arpp} (b). Due to the slightly fluctuating XUV flux, the Ar$^+$ yield varies slightly between the slices, and so does the Ar$^{2+}$ yield. The observed slope of $\sim 2$ in Fig.~\ref{fig:Arpp}(b) indicates that, indeed, two photons are involved in the ionization process. Since the ionization potential of Ar$^{+}$ (\SI{27.63}{eV}) is greater than the XUV photon energy, the underlying double ionization process cannot proceed via the ground state of Ar$^+$ assuming only contribution of H11. However, the presence of H13 in the experiment opens an additional sequential ionization pathway. The shares of the different pathways are discussed in detail in Ref.~\cite{Hell2025}.

For an estimation of the XUV peak intensity in the focus, we simulated the rates of generating Ar$^{+}$ and Ar$^{2+}$ using the absorption cross-section for one\cite{Samson2002} and two photons. The latter one was calculated from a recent measurement\cite{Hell2025}. The beam size at the focusing mirror $w = \SI{5.8}{mm}$ was calculated with the divergence (cf. Fig.~\ref{fig:scinty}). For an intensity of $I_0 = \SI{1.1+-0.5e11}{W/cm^2}$ and a focus size of $w_0 = \SI{1.4+-0.5}{\micro\meter}$ we could reproduce the measured rates. The resulting Rayleigh length of $z_R = \SI{18.3+-7.4}{\micro\meter}$ compared to the jet diameter $d_{jet} \approx \SI{440}{\micro\meter}$ indicates the presence of significant volume averaging along the propagation direction. Thus, the Ar$^{2+}/$Ar$^{+}$ rate could be increased by decreasing $d_{jet}$.

\subsection{XUV-VIS cross-correlation measurement}

In a second experiment, we perform a cross-correlation measurement of XUV and \SI{515}{nm} pulse. To this end, a part of the \SI{515}{nm} beam is sent into a second beam path including an interferometer to control the delay between XUV and \SI{515}{nm} pulses, cf.~Fig.~\ref{fig:XPETS_setup}. A perforated mirror with a \SI{4}{mm} hole is used to recombine the \SI{515}{nm} probe pulses back into the XUV beam path. Temporal and spatial overlap is achieved by temporarily removing the aluminum filter and employing the overlap control mirror to superimpose the \SI{515}{nm} probe with the residual \SI{515}{nm} for HHG generation collinearly.

\begin{figure}
\includegraphics[width=\linewidth]{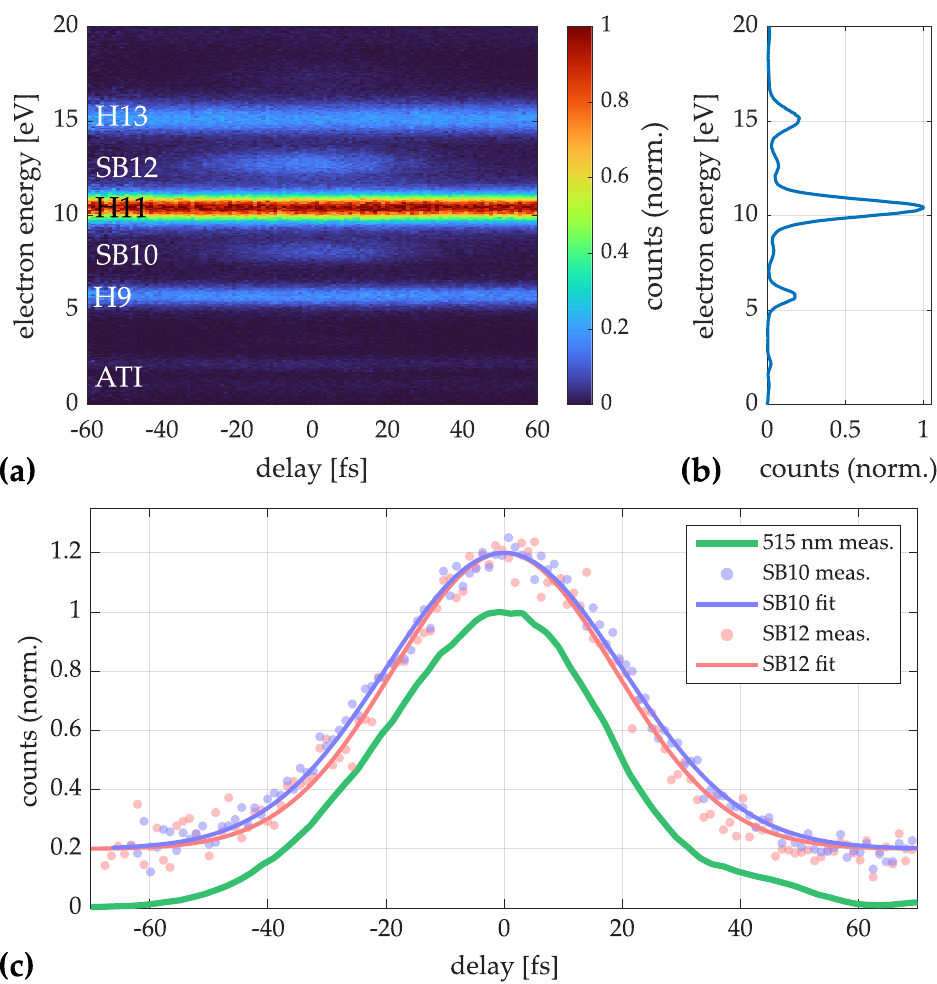}
\caption{\label{fig:XUV515} Cross-correlation experiment in argon using XUV and \SI{515}{nm} pulses. In (a), the photoelectron yield measured in coincidence with Ar$^+$ ions is shown as a function of electron energy and time delay between XUV and \SI{515}{nm} pulses. Negative delay indicates that the XUV pulse arrives first. SB and ATI denote the sidebands and above-threshold ionization, respectively. In (b), the normalized sum along the delay axis is shown. Panel (c) shows the measured \SI{515}{nm} pulse and the sidebands SB10 and SB12 (both shifted for visual convenience). To facilitate comparison, a Gaussian fit is added to each sideband.}
\end{figure}

First measurements were conducted using argon as a gas target, \SI{65}{\micro\joule} of \SI{515}{nm} light for HHG and \SI{4.4}{\micro\joule} for the probe beam.
Fig.~\ref{fig:XUV515} (a) shows the delay-dependent photoelectron spectra measured in coincidence with Ar$^+$ ions. The delay-independent photoelectron line at \SI{10.7}{eV}, corresponding to the absorption of H11, dominates the signal. In addition, weaker signals corresponding to photoionization by H9 and H13 are observed. Around the temporal overlap between XUV and \SI{515}{nm} pulses, additional sidebands, corresponding to the absorption or emission of additional \SI{515}{nm} photons, are observed. These originate from so-called laser-induced free-free transitions\cite{Glover1996,Schins1996}. The sub-cycle oscillations in the sidebands \cite{Paul2001} cannot be resolved in our experiment due to timing jitter between XUV and \SI{515}{nm} pulses.

The sidebands correspond to a cross-correlation between the XUV and \SI{515}{nm} pulses in the time domain, i.e., the observed sideband signal $I_\mathrm{SB}(t)$ is a convolution of the  \SI{515}{nm} pulse envelope $I_{515}(t)$ and the unknown XUV pulse $I_\mathrm{XUV}(t)$: 
\begin{equation}
    \label{eq:conv}
    I_\mathrm{SB}(t) = I_{515}^x(t)\ast I_\mathrm{XUV}(t),
\end{equation}
where the exponent $x$ may differ from 1 due to saturation of free-free transitions\cite{Bouhal1997}. 

In principle, the XUV pulse duration can be retrieved by deconvolution. The measured temporal intensity profiles of sidebands SB10 and SB12 are plotted in Fig.~\ref{fig:XUV515}(c). Also shown is the intensity profile of the \SI{515}{nm} probe pulse $I_{515}(t)$ in the COLTRIMS chamber. It was determined by adding the appropriate uncompensated spectral phase to the result of the FROG measurement\cite{Geib2019} presented in Fig.~\ref{fig:xuv_chamber} \footnote{The duration of the \SI{515}{nm} pulse used for HHG remains at $\approx\SI{30}{fs}$.}.  This results in a pulse duration (FWHM of the intensity envelope) of \SI{43.6 \pm 0.9}{fs}. For SB10 and SB12, we measure a FWHM of the temporal profile of \SI{47.4 \pm 1.6}{fs} and \SI{44.0 \pm 2.0}{fs}, respectively. The similarity of \SI{515}{nm} pulse duration and the FWHM of the sidebands supports that the XUV pulses are much shorter than the driving pulses. However, in this case, the deconvolution leads to extremely large uncertainties, such that an accurate retrieval of the XUV pulse duration is not possible based on the existing data.

A rough estimation of the XUV pulse duration is obtained based on the cut-off law of HHG and the intensity profile of the \SI{28}{fs} long \SI{515}{nm} pulse: the minimum intensity to generate H11 is $I_{\mathrm{min}} = \SI{1.4e14}{W/cm^2}$, while the peak intensity is \footnote{Calculated based on $P_L = \SI{6.5}{W}$ and the focus size determined in the appendix.} $I_0 = \SI{5+-1e14}{W/cm^2}$. By assuming a single-atomic scaling of the intensity of H11 as\cite{Klas2021a} $I_\text{H11} \propto I_0^{4.6}$ the XUV pulse duration is estimated to be \SI{11}{fs}.

\section{Conclusion and Outlook}

We have presented a new setup for coincidence spectroscopy with quasi-monochromatic XUV radiation based on HHG driven by short \SI{515}{nm} pulses at \SI{100}{kHz} repetition rate. The HHG source and beamline are equipped with various versatile diagnostic tools: optical imaging of the laser-induced plasma in the gas target; observation of the XUV beam by a scintillation crystal; live measurement of the XUV spectrum by a photoelectron time-of-flight spectrometer (XPETS). Using these tools, the efficiency of the HHG process was optimized by investigating in detail the effect of the gas nozzle position relative to the laser focus and the adjustment of the beam size using an iris aperture.\\
We observed maximum XUV flux for a nozzle position $\approx \SI{2}{mm}$ before the laser focus. Numerical calculations of the phase matching conditions confirm that good phase matching conditions are obtained before the focus for short trajectories. However, the predicted phase matching for long trajectories behind the focus is not observed experimentally. We further found that tuning the beam diameter with the iris can lead to a higher XUV flux caused by a higher microscopic response if generating out of the focus.\\
We reach an estimated photon flux of \SI{5e13}{photons/s} at \SI{26.5}{eV}, which corresponds to \SI{2.2}{nJ} pulse energy at a repetition rate of \SI{100}{kHz}. A specially coated concave mirror spectrally filters the XUV beam around \SI{26.5}{eV} and focuses it into the cold gas target of the COLTRIMS apparatus.\\
We showed first coincidence measurements of XUV photoionization of argon atoms. Owing to the high XUV flux and hard focusing, two-photon double ionization is achieved. This paves the way to study non-linear processes in the XUV regime with a lab-based setup. In particular, XUV pump and XUV probe coincidence experiments are within reach.\\
In addition, we demonstrate the capability to carry out pump-probe experiments by performing a cross-correlation of XUV and \SI{515}{nm} pulses. These measurements suggest that the XUV pulse duration is significantly shorter than the driving laser pulses. The XUV-pump \SI{515}{nm}-probe scheme will soon be used to study nuclear and electronic dynamics in a well-controlled manner.\\
Both schemes will benefit from a higher XUV flux. Towards this goal, the next steps are a further temporal compression of the \SI{515}{nm} pulses, since Klas et al. \cite{Klas2021a,Klas2021} demonstrated an increase in flux $\propto \tau^{-1}$, and the usage of krypton instead of argon, which may increase the flux up to a factor of five\cite{Klas2021a}. In addition, longer focusing on the gas target while maintaining the laser intensity could make it possible to use the remaining power of the \SI{515}{nm} to increase the XUV flux. 

\begin{acknowledgments}
We thank F. Ronneberger, Th. Weber and the FSU workshop for technical support. Helpful advice by F. Trinter, A. Czasch, S. Voss, Y. Mairesse and V. Blanchet is gratefully acknowledged. This project has been supported by the Deutsche Forschungsgemeinschaft (DFG, German Science Foundation) under the Emmy Noether programme project No. 437321733 and in the Collaborative Research Centre 1375 "Nonlinear optics down to atomic scales" (NOA) under projects B1 and Z3 (project No. 398816777). Further financial support has been provided by the profile line LIGHT by the Friedrich Schiller University and by the Max Planck School of Photonics.
\end{acknowledgments}

\section*{Author declarations}
\subsection*{Conflict of Interest}
The authors have no conflicts to disclose.

\section*{Data availability statement}
The data that support the findings of this study are available from the corresponding author upon reasonable request.

\section*{Appendix}

\subsection*{Focus profile}

\begin{figure}[h]
\includegraphics[width=\linewidth]{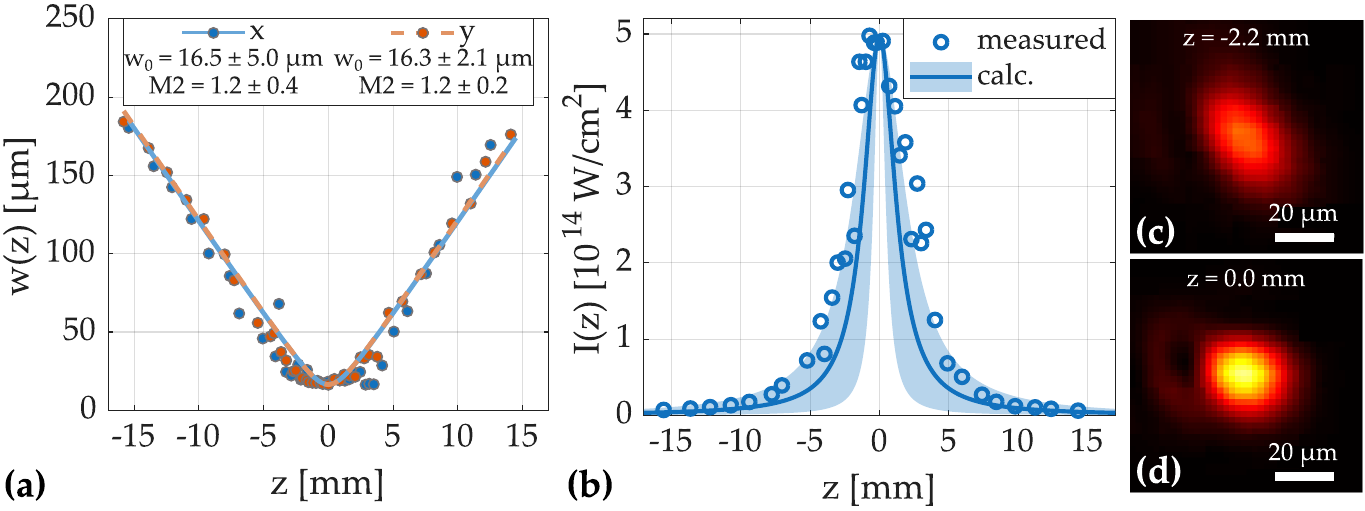}
\caption{\label{fig:focus} Beam radius and intensity along the focus for $f = \SI{200}{mm}$. In (a) the measured beam radius $w(z)$ (dots) and the fit to the Gaussian beam model (line) are shown for $x$ and $y$ direction. Panel (b) shows the intensity along the focus $I(z)$ derived from the measurement (dots) and derived from the Gaussian beam fit (line with error area). Further, images of the beam \SI{2.2}{mm} before (c) and at the focus (d) are presented.}
\end{figure}

The beam radius $w(z)$ along the focus was measured for the \SI{515}{nm} beam focused by a $f = \SI{200}{mm}$ lens, cf. Fig.~\ref{fig:focus} (a). This is the setup used for the experiments of Sec.~\ref{section:coincidence}. The focused beam is attenuated by the transmission through a \SI{515}{nm} HR mirror directly after the lens and several ND filters in front of the CMOS camera, which is used to capture images of the beam. Fitted to a Gaussian beam model\footnote{All following values are given as the mean of $x$ and $y$ direction.}, we obtain a focus size of $w_0 = \SI{16.4+-3.6}{\micro m}$ and a beam quality factor of $M^2 = \num{1.2+-0.3}$. This leads to a beam size at the lens of $w = \SI{2.4+-0.9}{mm}$ and a Rayleigh length of $z_R = \SI{1.4+-0.9}{mm}$ (mean value of $x$ and $y$ direction). The intensity along the focus $I(z)$ is presented in Fig.~\ref{fig:focus} (b) and extracted from the measurement assuming that the brightness of the images divided by the exposure time is proportional to $I(z)$. The related peak intensity for this measurement (average power $P_L = \SI{8.5}{W}$, pulse duration $\tau = \SI{38}{fs}$) is $I_0 = \SI{5.0+-2.4}{W/cm^2}$.

For the measurements presented in Sec.~\ref{section:HHG}, a lens with $f = \SI{150}{mm}$ was used. Taking the measured beam size at the lens of $w = \SI{4}{mm}$ and taking $M^2 = \num{1.2}$ from above, we obtain a focus size of $w_0 = \SI{7.4}{\micro m}$ and $z_R = \SI{277}{\micro m}$.

\section*{References}
\bibliography{hhg_chamber}

\end{document}